\documentclass{aa}  

\usepackage{graphicx}
\usepackage{txfonts}
\usepackage{xcolor}
\usepackage{hyperref}

\begin{document}

   \title{Modeling the circumstellar interaction around SN 2004gq }

   \subtitle{}

   \author{A. P. Nagy\inst{1},
           B. H. P\'al\inst{1} 
           \and
           T. Szalai \inst{1,2,3}
          }

   \institute{Department of Experimental Physics, Institute of Physics, University of Szeged, D{\'o}m t{\'e}r 9, 6720 Szeged, Hungary\\
    \email{nagyandi@titan.physx.u-szeged.hu}
    \and
     HUN-REN--SZTE Stellar Astrophysics Research Group, Szegedi {\'u}t, Kt. 766, 6500 Baja, Hungary
     \and
     MTA-ELTE Lend\"ulet "Momentum" Milky Way Research Group, Hungary
    }

   \date{}

 
  \abstract
   {The relationship between the mass-loss history and final evolutionary stage of massive stars and the properties of the observable supernova (SN) is still under debate. This is especially true for stripped-envelope (Type Ib/c) SNe, where the progenitor ejects a considerably large amount of material during its evolution, which can lead to a circumstellar medium relatively close to the exploding star. Moreover, when the star explodes as a SN, this matter may contribute significantly to the generated luminosity because of the interaction. However, the trace of this circumstellar interaction can only be investigated for a couple of Type Ib/c SNe, and the nature of a close (within around $10^{15}$ cm) circumstellar matter (CSM) has also been largely unexplored for these objects.}
   {We present the results of our radio and bolometric light curve (LC) analysis related to SN 2004gq. We describe a combined model that explains the unusual LC properties of this event and supports the circumstellar interaction scenario.}
   {We computed the quasi-bolometric LC of the SN and fit this with a multicomponent model to gain information on the progenitor and the surrounding circumstellar medium. We also analyzed the available radio LCs (taken at 1.4,\ 4.9 and 8.5 GHz) of SN 2004gq to verify our estimated average mass-loss rate, which is one of the most crucial physical properties related to CSM models.}
   {We infer reasonable parameters for SN 2004gq using radioactive decay and magnetar energy input.  To power the entire LC, we must also add an extra energy source related to the CSM. We determine the most essential parameter of this medium: the average mass-loss rate from both LC and radio data fitting. We find that the suggested hidden circumstellar interaction is a viable mechanism that provides the required energy deficiency and that it can be estimated using a simple semi-analytic model.}
   {}

   \keywords{Methods: analytical --
                (Stars:) supernovae: individual: SN 20004gq --
                (Stars:) circumstellar matter
               }

   \maketitle
%

\section{Introduction}

Type Ib and Ic supernovae (SNe) form a unique group among core-collapse SNe  that show no hydrogen and/or no helium in their optical spectra. The classification of SNe can change over time \citep[e.g.,][]{Modjaz, Milisavljevic}, and thus we usually refer to this group as Type Ib/c or stripped-envelope SNe \citep[SESNe,][]{Clocchiatti}. Conversely, it is also possible that a normal Type Ic SN shows similar spectroscopic features as interacting Type Ibn and Icn SNe at late phases \citep[e.g.,][]{Kuncarayakti}.

While we are aware of a significant mass loss during the pre-SN evolution, the exact mechanism at play is under debate. Some observations suggest \citep[e.g.,][]{Cao} a possible Wolf-Rayet progenitor that loses its outer envelope due to stellar wind. However, other studies \citep[e.g.,][]{Sana, Woosley} assume a binary interaction prior to the explosion that stripped away the outermost layers of the massive star. These phenomena suggest the presence of circumstellar matter (CSM) around the progenitor regardless of the exact mechanism. Thus, studying the nature of this CSM could play an important role in revealing the progenitor's mass-loss history just before the explosion. 

Observing radio emissions from SNe offers a particular advantage in this field: it traces the fastest ejecta, which are difficult or impossible to detect optically \citep{Bietenholz}. Moreover, as the ejecta interact with their surroundings, a forward and a reverse shocks are formed. The movements of these shock fronts generate synchrotron emission, which is created exclusively from the SN-CSM interaction \citep[e.g.,][]{Chevalier, Maeda, Matsuoka}. Moreover, as \cite{Moriya1} reveal, the rise time and peak luminosity in radio are strongly related to the CSM density, and therefore to the mass-loss history of the progenitors. As such, the rise time and peak luminosity increase with higher CSM density.

Previous studies \citep{Yaron, Maeda21} show that the typical CSM radius for SESNe is $\sim 10^{15}$ cm, which suggests that this matter should be ejected just a few months prior to the SN explosion. However, assuming a high-velocity wind for this mass-loss event, the expected CSM density would not be sufficiently high to leave a solid optical contribution. Nevertheless, a few examples (e.g., SNe 2001em, 2004dk, and 2014C) exist where the interaction between the SESN ejecta and the CSM produces strong emissions in optical and/or radio wavelengths \citep[e.g.,][]{Pooley_2019, Margutti}. 
However, in these cases the interaction occurred several months or years after the explosion, which means that the CSM is located too far from the progenitor to be created by an extensive mass-loss event within the last few years of the exploding star.  

 Although recent studies reveal a possible connection between CSM interaction and a late-time (200$\lesssim$ days after the explosion) re-brightening of SESNe \citep{Sollerman,Soraisam,Kuncarayakti1}, they do not consider smaller, earlier light curve (LC) bumps occurring around 60-100 days after the explosion. Thus, a direct connection between CSM interaction and these bumps has not been shown, but we assume that such a connection could exist. We predict that this excess luminosity arises from an additional power source related to the interaction between the SN ejecta and a close CSM \citep[e.g.,][]{Chevalier}, as the mass-loss history shortly before the SN explosion can drastically influence the optical LC properties (e.g., brightness and color) of Type Ib/c progenitors \citep{Jung}. To test this theory, we chose to examine the mostly neglected Type Ib/c SN 2004gq, which has detailed and well-sampled data in both the optical and radio. 
 
The paper is organized as follows: In Section 2 we present the dataset of SN 2004gq. In Section 3 we describe the basis of our models related to the quasi-bolometric LC fitting. We introduce the radio data analysis in Section 4. Section 5 presents our calculations associated with the average pre-SN mass-loss rates. Finally, in Section 6 we summarize our conclusions.

\section{Dataset}

SN 2004gq was discovered on December 11, 2004 (approximately 2004.36 UT) in galaxy NGC 1832 \citep{iauc}. The explosion time of this SN is well constrained (December 7 $\pm$ 2 days, 2004) because of a non-detection just 6 days before the first observation. SN 2004gq was originally classified as a Type Ic SN, but was later reclassified as Type Ib due to its later-phase ($\sim$30 days) spectrum showing strong helium lines. This phenomenon may suggest a later interaction with a helium-rich CSM. 

Very Large Telescope (VLT) radio observations of SN 2004gq were obtained between December 16, 2004, and March 26, 2006, at four different frequencies ($\nu = 1.4,\, 4.9,\, 8.5,$ and $15$ GHz) by \citet{Wellons}. For our own radio analysis, we used the integrated flux densities from their paper at 1.4 GHz, 4.9 GHz, and 8.5 GHz between 8 and 300 days after the explosion. We set aside the 15 GHz frequency because of insufficient observational data for radio LC modeling. It was not possible to model the radio LC around the peak flux densities.

We also used optical and near-IR observations to build the quasi-bolometric LC of SN 2004gq, which was required to compare the measurements with models. We first collected the measured magnitudes in all available photometric ($UBVRJH$ and $ugri$) bands from the Open Supernova Catalog\footnote{currently at  \href{https://github.com/astrocatalogs}{https://github.com/astrocatalogs}}. 
 The optical photometry included all of the $BVugri$, $UBVI$, and $VR$ data obtained with the Swope Telescope \citep{Stritzinger}, the Fred L. Whipple Observatory (FLWO) 1.2 m telescope \citep[CfA3 campaign,][]{Bianco}, and the Palomar 60-inch telescope \citep{Drout}, respectively. The near-IR photometry was measured in J and H bands with the FLWO Peters Automated Infrared Imaging Telescope (PAIRITEL) and the Irénée du Pont Telescope \cite{Bianco}.

These data were then converted into fluxes using extinctions, distance, and zero points \citep{Bessell}. The Galactic extinction values for the individual bands were taken from the NASA/IPAC Extragalactic Database (NED). To be consistent with the radio data, we used 26 Mpc as the distance of the SN. Photometric measurements were not equally sampled across all filters. Thus, at epochs when the data were missing, we linearly interpolated the flux from nearby values. The trapezoidal rule was applied while integrating over all wavelengths, with the assumption that flux reaches zero at 2000 $\mathring{A}$. The IR contribution was taken into account by the exact integration of the Rayleigh-Jeans tail, from the wavelength of the last available photometric near-IR band to infinity.
 
\section{Light curve interpretation}
The quasi-bolometric LC of SN 2004gq is unusual for a core-collapse SN as it does not follow the nickel-cobalt decay rate at late times, even when gamma-ray leakage is taken into consideration \citep{Wheeler}. The initial, steady luminosity decline of the LC tail appears to re-brighten slightly after 60 days. This phenomenon could suggest a CSM shell around the exploding star at a distance of $10^{15} - 10^{16}$cm, corresponding to a recent mass-loss event. 

To convert these qualitative LC features into quantitative considerations, we assumed that the observable light variations in different bands, and consequently the quasi-bolometric LC, were generated by three different energy inputs: photo-diffusion, magnetar spin-down, and CSM interaction. 
In our estimated progenitor configuration, we had spherically symmetric ejecta and a CSM shell. While both regions have a common center, they are separated from each other. Hence, the generated shock wave requires some time to reach the CSM. This configuration offers two advantages: it ensures self-consistency with the radio data and general LC properties, and it allows the differential equations of the two components to be separated, as the photon diffusion timescale is much shorter in one of the regions \citep{Kumar}. Thus, the generated luminosities at late times could be simply summed.

To determine the major physical properties of the SN explosion, we used the semi-analytical model originally presented by \citet{Arnett} and later generalized by \citet{Nagy1}. This model assumes homologous expanding SN ejecta, where the energy loss is driven by radiation transport and is treated by diffusion approximation. The bolometric luminosity is determined by two processes: the radioactive heating from nickel and cobalt decay and the energy released from the spin-down of a newly born magnetar. To ensure self-consistency with the supposedly low ejected mass of the SN, we took into account the gamma-ray trapping as 
\begin{equation}
    L_{bol} = L_{sn}\ (1 - exp(-T_0^2/t^2)),
\end{equation}
where $L_{sn}$ is the energy released from radioactive heating and magnetar energy input \citep[][]{Arnett, Nagy} and $T_0$ is the characteristic timescale of gamma-ray leakage. 

A detailed description of the applied equations, modeling properties, and parameter correlations can be found in \cite{Nagy} and \cite{Nagy1}.  The most crucial model parameters we attempted to determine were the initial radius of the progenitor ($R_0$), the initial kinetic energy ($E_{kin}$), the ejected mass ($M_{ej}$), the generated nickel mass ($M_{Ni}$), the initial rotational energy of the magnetar ($E_p$), and the characteristic timescale of the spin-down ($t_p$). For the specific case of SN 2004gq, the best-fit parameter values for the SN ejecta were as follows: $M_{ej} = 1.6\, M_\odot$, $M_{Ni} = 0.048\, M_\odot$, $R_0 = 4.1 \times 10^{11}$ cm, $E_{kin} = 1.2 \times 10^{51}$ erg, $E_p = 1.7 \times 10^{48}$ erg, and  $t_p = 7$ days. These parameters indicate an average expansion velocity of $\sim 11 200$ km/s. 

While the average expansion velocity should be around this value, the ejecta mass and the explosion energy can only be constrained within a factor of two due to parameter correlation. We also acknowledge that the explosion energies from our code are usually higher than those obtained from more complex hydrodynamic models \citep[see the more detailed discussion in][]{Nagy1}.

 \cite{Barbarino} examined the evolution of the bolometric LC for 41 Type Ic SNe and found significant diversities in late-time LC decay rates and physical properties. The ejecta mass, the kinetic energy, and the ejected nickel mass of this sample varied from 0.12-31.29 $M_\odot$, (0.08 - 10.86)$\times 10^{51}$ erg, and 0.04-0.63 $M_\odot$, respectively. They also obtained average values of these parameters as $\langle M_{ej}\rangle= 4.39 \pm 0.31\, M_\odot$, $\langle E_{kin}\rangle = (1.71 \pm 0.15)\, \times 10^{51}$ erg, and $\langle M_{Ni}\rangle = 0.19 \pm 0.05\, M_\odot$. These results indicate that SN 2004gq lies at the lower end of this group. In contrast, SN 2004gq was modeled using two different approaches by \citet{Taddia1}: a semi-analytical calculation and a one-dimensional hydrodynamic model. These revealed that the ejecta mass, kinetic energy, and nickel mass should be 2.5 $\pm$ 1.5 $M_\odot$, (2.0 $\pm$ 1.3)$\times 10^{51}$ erg, and 0.08 $\pm$ 0.02 $M_\odot$, respectively. Our predicted values fall mostly within these parameter ranges. The only exception is the nickel mass, for which we obtained a lower value. However, this effect may be due to us using the magnetar energy input in addition to the radioactive decay, as this extra energy source affects the LC tail, which decreases the required nickel masses. Overall, our results show reasonable agreement with previous studies.    

To estimate the basic physical properties of the underlying CSM, we used an analytic model published by \citet{Chevalier1}. In this scenario, expanding SN ejecta collide with a nearly stationary CSM, resulting in a reverse and a forward shock wave that generates an extra energy source through the thermalization of kinetic energy.  

Assuming a thin shocked shell with an $M_s$ swept-up mass and $R_s$ radius, we calculated the momentum equation as 
\begin{equation}
 M_s\ \frac{d^2R_s}{dt^2}=4\ \pi\  R^2_s\ \left[\rho_{ej}\ \left(\frac{R_s}{t}-\frac{dR_s}{dt}\right)^2- \rho_{cs}\ \left(\frac{dR_s}{dt}\right)^2\right]\ ,
\end{equation}
where $\rho_{ej}$ and $\rho_{cs}$ are the density of the ejecta and the CSM, respectively. 

The swept-up mass was assumed to be the sum of the mass behind the reverse shock ($M_{rev}$) and the mass of the shocked CSM ($M_{cs}$). If the density profile exponent of the ejecta is $n$, the mass behind the reverse shock can be calculated as
\begin{equation}
M_{rev} = \frac{4\, \pi\, \rho_0\, t^3_{int}\ v^n_0\ t^{n-3}}{(n-3)\ R^{n-3}_s}\  ,
\end{equation}
where $\rho_{0}$ and $v_0$ are the density and velocity of the ejecta at the starting time of the interaction ($t_{int}$), which should be around 60-90 days, according to the LC. Assuming that the density profile index of the CSM ($\gamma$) is arbitrary, indicating that it was not created by a steady wind, the mass of the shocked CSM is
\begin{equation}
M_{cs} = \frac{\dot{M}\ R^{3-\gamma}_s}{(3-\gamma)\ w}\ ,
\end{equation}
where $\dot{M}$ is the average mass-loss rate and $w$ is the velocity of the stellar wind. Here, we assumed a Wolf-Rayet star as a progenitor, characterized by a fast stellar wind with an estimated velocity between 1000 and 2500 km/s \citep{Chevalier1}. 

Solving these equations enabled the calculation of $R_s$ at every time step, which allowed us to determine the luminosity generated by the CSM interaction:
\begin{equation}
 L_{CSM}= 2\ \pi \ \rho_0\ t^3_{int}\ v^n_0\ t^{n-6}\ R^{5-n}_s\ \left(\frac{3-\gamma}{n-\gamma}\right)^3\ .
\end{equation}

\begin{figure}[!h]
\centering
\includegraphics[width=0.48\textwidth]{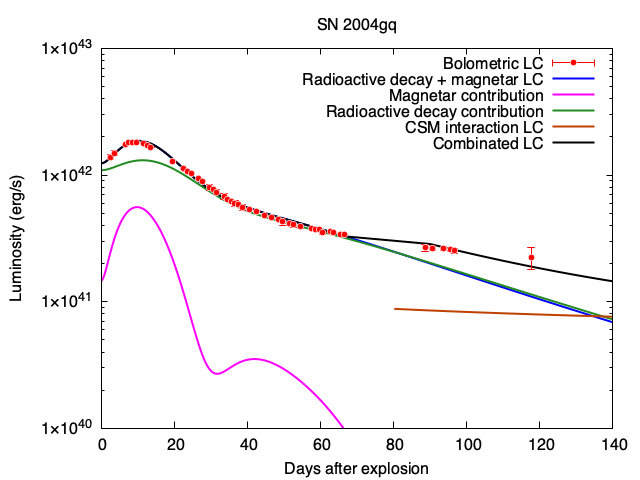}
\caption{Comparison of the quasi-bolometric LC of SN 2004gq (red dots) with best-fit model data. The orange and blue curves represent the contribution of the CSM interaction and the  photon-diffusion model, respectively, while the black lines show the combined LCs.}
\label{fig:lc}
\end{figure}

To fit the entire quasi-bolometric LC of SN 2004gq, we summed the luminosities derived from the SN explosion model and those from the CSM interaction. Thus, we should be taking all the energy production processes after the start of the interaction into consideration. Our final results suggest that $t_{int}$ was more likely around 80 days, which is in line with the general LC properties (Fig. \ref{fig:lc}). The derived physical properties of the circumstellar shell were $w = 1300$ km/s, $\dot{M} = 3.5 \times 10^{-3}$ $M_\odot$/yr, $\gamma = 1.78,$ and $n = 10$, which correspond well to a recently formed CSM. 

\section{Radio analyses}
In contrast to optical or near-IR observations, the radio signal can trace the outer, faster-moving layer of the SN explosion. Moreover, as the shock wave interacts with the CSM, it generates synchrotron radiation, which can be detected at radio wavelengths.

To determine the average mass-loss rate that generated the CSM, we fit the radio light curves. One possible way to approach this was to use the dynamical model to determine the physical properties of the CSM around SN 2004gq as \cite{Wellons} previously demonstrated. However, their assumption regarding a characteristic timescale for synchrotron self-absorption (SSA) of 10 days appears to be relatively short compared to the maxima of the radio LCs at a given frequency.  Furthermore, another widely used radio model in the literature \citep{Weiler1,Weiler} provides a more detailed description of the CSM.

For this analysis, we used the parametric model published by \citet{Weiler1, Weiler}, which calculates the spectral flux density in mJy as
\begin{equation}
 S_\nu = K_1\left(\frac{\nu}{5 \ GHz}\right)^\alpha \left(\frac{t-t_0}{1 \ day}\right)^\beta  \left(\frac{1-e^{-\tau_{CSM,cl}}}{\tau_{CSM,cl}}\right) \left(\frac{1-e^{-\tau_{int}}}{\tau_{int}}\right)e^{-\tau_{ext}} ,   
\end{equation}
where $K_1$ denotes the dimensionless flux density and $t_0$ is the time of the explosion. The different $\tau$ parameters represent the various attenuation factors (or optical depths) of electromagnetic radiation: $\tau_{ext}$ and $\tau_{int}$ refer to external and internal absorption, respectively. 

The external optical depth originally contains two homogeneous free-free absorption (FFA) processes \citep{Weiler}. Given the available early radio dataset, we neglected the distant component, as this effect would become more significant at later times and lower frequencies. As such, in our analyses $\tau_{ext} \approx \tau_{CSM_{hom}}$ and was calculated as
\begin{equation}
\tau_\textrm{CSM,hom} = K_2 \left(\frac{\nu}{5 \ GHz}\right)^{-2.1} \left(\frac{t-t_0}{1 \ \textrm{day}}\right)^{\delta},
\end{equation}
where $K_2$ corresponds to the dimensionless flux density of the homogeneous FFA.

In contrast, the internal attenuation factor includes SSA and thermal emission. 
While \citet{Chevalier} demonstrated that SSA is the dominant process for Type Ib/c SNe, nonthermal SSA is more significant at early times \citep{Chevalier_1998}. Furthermore, the FFA component contributes increasingly as the emission becomes more optically thin \citep{Weiler2}. Additionally, in certain cases like Type Ib SN1998bw \citep{Weiler1} and Type Ic SN1994I \citep{Weiler2}, thermal absorption was not negligible. Therefore, the internal absorbent region consists of two parts, an internal SSA and an internal FFA, which are as follows: 
\begin{equation}
\tau_{int_{SSA}} = K_5 \left(\frac{\nu}{5 \ GHz}\right)^{\alpha-2.5} \left(\frac{t-t_0}{1 \ \textrm{day}}\right)^{\delta''},
\end{equation}
where K$_5$ corresponds to the dimensionless flux density of the internal, nonthermal SSA; and

\begin{equation}
\tau_{int_{FFA}} = K_6 \left(\frac{\nu}{5 \ GHz}\right)^{-2.1} \left(\frac{t-t_0}{1 \ \textrm{day}}\right)^{\delta'''},
\end{equation}
where K$_6$ corresponds to the dimensionless flux density of the internal, thermal FFA.
   
The $\tau_\textrm{CSM, cl}$ describes the attenuation factor for a local clumpy or filamentary free-free absorbing CSM. The CSM should be relatively close to the SN progenitor star. In this context, "relatively close" means that the CSM is affected by the rapidly expanding SN shock wave within 100 days, precisely as the bolometric LC analysis suggests. To determine the exact value of this parameter, we used the following equation:   
\begin{equation}
\tau_\textrm{CSM,cl} = K_3 \left(\frac{\nu}{5 \ GHz}\right)^{-2.1} \left(\frac{t-t_0}{1 \ \textrm{day}}\right)^{\delta'} \ ,
\end{equation}
where $K_3$ refers to the dimensionless flux density of the clumpy or filamentary FFA.

\begin{table*}[!h]
\begin{center}
\caption{Best-fit model parameters for the FFA radio LC model}
\begin{tabular}{lcccccccc}
\hline
\hline
\noalign{\smallskip}
 & $\beta$ & $\delta$ & $\delta'$ & $\delta'''$ &$K_1$ & $K_2$ & $K_3$ & $K_6$\\
\noalign{\smallskip}
\hline
\noalign{\smallskip}
 1.4 GHz & -0.46 & -2.28 & -3.95 & -2.22 & 2.20$\ \times 10^{-6}$ & 20.03 & 3634.39 & 1732.44 \\
\noalign{\smallskip}
4.9 GHz & -1.95  & -0.46 & -2.61 & -1.97 & 1.07$\ \times 10^{-5}$ & 12.42 & 4852.78 & 342.46 \\
\noalign{\smallskip}
8.5 GHz & -1.54 & -2.04 & -2.08 & -1.27 & 6.02$\ \times 10^{-4}$ & 5.12 & 2126.23 & 173.02\\
\noalign{\smallskip}
\hline
\end{tabular}
\label{tab:radio_modelfit_FFA}
\end{center}
\end{table*}

\begin{table}[!h]
\begin{center}
\caption{Best-fit model parameters for the SSA radio LC model}
\begin{tabular}{lcccc}
\hline
\hline
\noalign{\smallskip}
& $\beta$& $\delta''$ &$K_1$ & $K_5$ \\
\noalign{\smallskip}
\hline
\noalign{\smallskip}
 1.4 GHz & -0.44 & -2.39 & 8.44 & 596.87 \\
\noalign{\smallskip}
 4.9 GHz & -1.00  & -3.99 & 226.15 & 604255.0 \\
\noalign{\smallskip}
8.5 GHz & -1.42 & -3.09 & 1546.65 & 111339.0  \\
\hline
\end{tabular}
\label{tab:radio_modelfit}
\end{center}
\end{table}

During the analysis, we fit the available radio data with the previously described equations using a $\chi^2$  minimization method. To reduce the parameter space, we tried to apply certain restrictions for the parameters. However, according to previous studies \citep[e.g.,][]{Chevalier, Weiler2}, only the radio spectral index ($\alpha$) can be restrained, as it should be somewhere between $-$2.5 and $-$1 for Type Ib/c SNe. Following the work of \cite{Weiler2}, we calculated the spectral index parameter, which was around $\alpha\approx-1.0$. We used this value for both FFA and SSA modeling. For all other parameters, we set an initial value for the fit with a similar order of magnitude as published by \citet{Weiler1}. 
 
\begin{figure}[!h]
\centering
\includegraphics[width=0.48\textwidth]{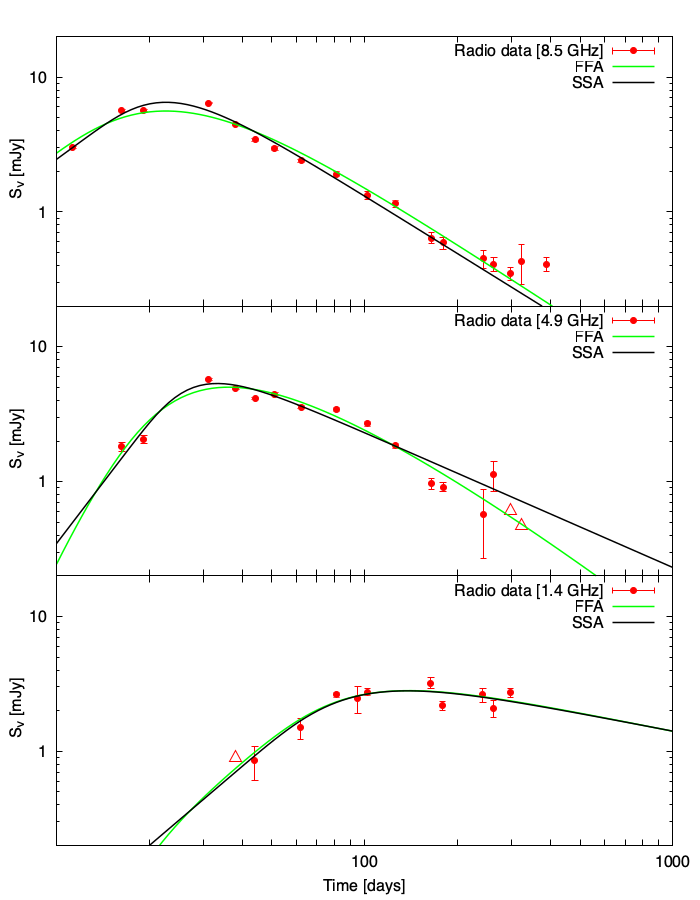}
\caption{Radio LCs of SN 2004gq at 8.5 GHz (top panel), at 4.9 GHz (middle panel), and at 1.4 GHz (bottom panel). The red dots show the observational data at the specified frequency, and the red triangles show the upper limits of the observations, which were not included in the model fits. The green lines represent the best-fit ‘‘pure’’ thermal FFA model, and the black lines represent the best-fit "pure" nonthermal SSA.}
\label{fig:radio}
\end{figure}

As a result, we obtained minimal $\chi^2$ values for the FFA and SSA  best-fit modeling as follows: $\chi^2_{FFA}=0.99, \ 2,30, \ 2.32,$ at 1.4, 4.9, and 8.5 GHz and $\chi^2_{SSA}=0.98, \ 1,93, \ 1.49,$ at 1.4, 4.9, and 8.5 GHz, respectively. We also estimated the uncertainty of these calculations by fitting the errors of the flux density. These values were consistent with those published by \citet{Weiler2}. 
The relevant parameter values for this best-fit model can be seen in Table \ref{tab:radio_modelfit}. In addition, Fig. \ref{fig:radio} illustrates the radio data of SN 2004gq and the generated model curves created by the calculated quantities.

\section{Mass-loss rate estimations}
The average mass-loss rate is the most crucial parameter related to the pre-SN evolution and CSM generation of the progenitor star. Our combined model can calculate $\dot{M}$ naturally from the fit of the quasi-bolometric LC. However, according to the parameter correlations, the estimated mass-loss rate value should only be considered as an order of magnitude estimation. Thus, we aimed to verify this approximated value by determining the average mass-loss rates from the radio data fit and comparing them to the gained values.

It is possible to estimate the average mass-loss rate of the SN progenitor from the fitted radio absorption parameters \citep{Stockdale} as
\begin{eqnarray}
\label{eqn:abs_mass}
\dot{M} &=& 3.0 \times 10^{-6}\ \left( \frac{ t^m}{t_{i} ^{m}\ m} \right )^{3/2} \left(\frac{t_i}{45\, \textrm{days}}\right)^{3/2}\left(\frac{\overline{v}}{10^4\, \textrm{km/s}}\right)^{3/2} \\ \nonumber & & \times  \left(\frac{T}{10^4\,  \textrm{K}}\right)^{0.68} \ \left(\frac{w}{10\, \textrm{km/s}}\right)\ \tau^{1/2}_{eff}\Phi \,
,\end{eqnarray}
where $T = 100\, 000$ K is the electron temperature \citep{Chevalier17} in the stellar wind, $m$ is a modeling constant \citep{Weiler}, $t_{i} \approx 45$ days \citep[e.g.,][]{Weiler1, Weiler}, $\overline{v}$ is the average shock velocity, $w$ is the wind velocity, $\tau_{eff}$ is the effective optical depth of the CSM, and $t\approx1$ is the time of the shock breakout \citep{Weiler}. 
Using the same density profile as derived from the bolometric LC fit ($\rho_{CSM}\sim r^{-1.78}$), the $\Phi$ factor will be $0.92$ \citep{Weiler}.
The effective optical depth of the CSM can also be calculated from the fitted radio parameters as
\begin{eqnarray}
\tau_\textrm{eff}^{1/2} = 0.67\times\frac{(\tau_\textrm{CSM,hom}+\tau_\textrm{CSM,cl})^{3/2}-\tau_\textrm{CSM,hom}^{3/2}}{\tau_\textrm{CSM,cl}} .
\end{eqnarray}

To ensure self-consistency in this comparison, we used the wind velocity gained from the quasi-bolometric LC fit with a value of 1300 km/s. However, due to the lack of spectroscopic data, the average shock velocity at a given frequency must be estimated using the following expression \citep{Milisavljevic, Chevalier}:

\begin{equation}
    \overline{v} \approx 0.13\  c\ \left(\frac{5\, \textrm{GHz}}{\nu}\right) \ \left(\frac{10\, \textrm{days}}{t_p}\right)\ \left(\frac{\epsilon_l}{\epsilon_B}\right)^{-1/19}\  \left(\frac{d}{10\, \textrm{Mpc}}\right)^{18/19} \  S_{p}^{9/19} , 
\end{equation}
 where $\epsilon_l$ and $\epsilon_B$ describe the efficiency of the shock wave in accelerating electrons and amplifying magnetic fields \textbf{( $\epsilon_l = \epsilon_B = 1/3$)}. As the radio data suggest (Fig. \ref{fig:radio}), at 1.4 GHz $t_p\approx$ 164 days and $S_p\approx$ 3.21 mJy and at 4.9 GHz $t_p\approx$ 31 days and $S_{p}\approx$ 5.75 mJy, while at 8.5 GHz  $t_p\approx$ 21 days and $S_{p}\approx$ 6.4 mJy. Hence, the average shock velocities for 1.4 GHz, 4.9, and 8.5 GHz were $3.44 \times 10^4$, $6.86 \times 10^4$, and $6.14 \times 10^4$ km/s, respectively.

Furthermore, the average mass-loss rate of the progenitor can be directly estimated from parameters related to the radio emission as

\begin{equation}
\dot{M} = 8.6\times 10^{-9} \ \left(\frac{L_\textrm{p}}{10^{26} \ \textrm{erg/s/Hz}}\right)^{0.71}\, \left(\frac{w}{10 \ \textrm{km/s}}\right)\  t_\textrm{p}^{1.14}\ ,
\end{equation}
where $L_\textrm{p}$ is the peak luminosity of the radio LC at a given frequency. Here, the $L_\textrm{p}$ were $2.53 \times 10^{27}$, $4.53 \times 10^{27}$,  and $5.50 \times 10^{27}$ erg/s/Hz at 1.4, 4.9, and 8.5 GHz, respectively. Although this calculation usually gives reasonable mass-loss rates \citep{Weiler1}, its limitations should be taken into account. Specifically, this equation assumes similar characteristics for all radio SNe and relies solely on luminosities at 5 GHz. Consequently, our calculated values are plausible only within an order of magnitude.

In conclusion, these different calculations for the average mass-loss rate suggest a similar parameter range (Table \ref{tab:mass-loss}), and the $\dot{M} =3.5 \times 10^{-3}\ \textrm{M}_\odot/\textrm{yr}$ we determined from the quasi-bolometric LC fit is in good agreement with these values. 
Furthermore, it was previously demonstrated \citep[e.g.,][]{Moriya, Yaron} that a CSM created by pre-SN activity in the last $\sim$ 30 years should correspond to a mass-loss rate of $\sim 10^{-3} - 10^{-4}\,  M_\odot/yr$, which also confirms our results. 

We note that Eq. \eqref{eqn:abs_mass} depends on several constant parameters. For instance, if we assume that the progenitor was a red supergiant with a significantly lower wind velocity ($w = 10 \ km \ s^{-1} $), the mass-loss rate values would change by two orders of magnitude ($\dot{M}\sim 10^{-5} \ M_{\odot}/yr$). Nevertheless, our results are significantly higher for the same miscellaneous parameters compared with those described in \cite{Wellons}. This discrepancy could arise from the inclusion of FFA processes, which alter the initial conditions of the modeling and result in different fitting parameter values. As \cite{Soderberg_2005} showed for the Type Ic SN SN2003L, using a combined SSA and FFA model results in a higher value for average mass-loss rates compared to applying the SSA process alone, as shown by  \cite{Wellons}. Therefore, it seems reasonable to gain higher values if we implement FFA and SSA in our calculations. Moreover,  our result is on the same order of magnitude as mass-loss rates estimated from radio absorption and optical modeling.

\begin{table}[!h]
\begin{center}
\caption{Average mass-loss rate values estimated from radio data}
\begin{tabular}{lcc}
\hline
\hline
\noalign{\smallskip}
 &  absorption & emission\\
 \noalign{\smallskip}
\hline
\noalign{\smallskip}
$\dot{M}_{1.4}$ (M$_\odot$/yr) & (1.09 $\pm^{0.026}_{0.002})\times 10^{-3}$ & (2.90 $\pm^{0.189}_{0.185})\times 10^{-3}$\\
 \noalign{\smallskip}
$\dot{M}_{4.9}$ (M$_\odot$/yr) & (1.70 $\pm^{0.021}_{0.130})\times 10^{-3}$ & (4.39 $\pm^{0.004}_{0.004})\times 10^{-3}$\\
 \noalign{\smallskip}
$\dot{M}_{8.5}$ (M$_\odot$/yr) & (8.78 $\pm^{0.002}_{0.042})\times 10^{-4}$ & (4.74 $\pm^{0.003}_{0.003})\times 10^{-3}$ \\ 
\noalign{\smallskip}
\hline
\end{tabular}
\label{tab:mass-loss}
\end{center}
\end{table}

\section{Conclusions}
The mass of synthesized nickel is one of the most important power sources for SNe because it helps constrain the explosion mechanisms of these objects. However, for Type Ib/c SNe, it is a well-known fact that the estimated nickel masses are systematically higher than those estimated for Type II SNe. However, our results do not support this phenomenon, which may be due to the different (nickel decay + magnetar energy input) explosion mechanism assumed in our calculations. Recent studies \citep{Sollerman1, Ouchi, Meza} suggest that considering observational biases may reduce the mean nickel mass value for SESNe. For example, \citet{Sollerman1} determined that  $M_{Ni}$ should be around 0.16 $M_.\odot$. However, according to \cite{Meza} and \cite{Ouchi}, the mean nickel masses for Type Ib and Ic SNe at lower distances (d < 100 Mpc) are about 0.08 $M_\odot$ and 0.05 $M_\odot$, respectively. Our estimated Ni mass is slightly lower than these values, which could be the effect of the previously mentioned energy input scenario or the parameter correlations within our model. Thus, we can assert that our calculations are consistent with the predicted nickel mass for a nearby Type Ib/c SN. Alternatively, recent studies \citep[e.g.,][]{Ergon} indicate that, for Type IIb SNe, nickel bubbling could have a significant effect on the LC tail. Considering this scenario, nickel mixing might explain the observed luminosity excess of SN 2004qg at late times. However, only a higher nickel mass would result in sufficient mixing to account for the observed extra luminosity in the late-time  LCs. For other Type Ib/c SNe, we should consider whether nickel bubbling, CSM interaction, or other physical effects may cause a late-time re-brightening of the SN LC.

The radio analysis and the analytical LC modeling yield consistent conclusions regarding the average mass-loss rate. Furthermore, the calculated distance of the CSM is also self-consistent. Specifically, the radio data suggest a minimal CSM radius of around $8.0 \times 10^{15} - 1.95 \times 10^{16}$ cm, while the value calculated from the collision time (80 days) and average expansion velocity is $7.75 \times 10^{15}$ cm.    

The previously described model may be useful for generating the entire light variation, shortly after shock breakout throughout the nebular phase, of Type Ib/c SNe showing bumpy LCs. In addition, the reasonably good agreement of the radio and optical analyses also suggests that our modeling configuration could be a useful tool for preliminary studies, capable of providing quick estimates for essential parameters of SNe as well as its CSM interaction. This method could be particularly valuable for Type Ib/c SNe that lack spectroscopic or radio observations. 

We note that our suggestion of a hidden CSM interaction, which cannot be seen in the spectra as narrow features, is only plausible if the CSM is relatively close to the progenitor and settled in a disk or torus configuration. Nevertheless, our analytic calculations are only capable of modeling a spherically symmetric environment surrounding the exploding star. Despite this limitation, the overall modeling outcome seems reasonable. However, we should be aware that these results are only used for order-of-magnitude estimations due to the basic features of semi-analytical models. Thus, to reveal the real nature of Type Ib/c SNe, these estimated fitting parameters could be used for preliminary studies before applying more sophisticated hydrodynamic calculations.

\begin{acknowledgements}
This project is supported by NKFIH/OTKA PD-134434 and FK-134432 grants of the National Research, Development and Innovation (NRDI) Office of Hungary.
T.S. is supported by the János Bolyai Research Scholarship of the Hungarian Academy of Sciences, and by the New National Excellence Program (UNKP-22-5) of the Ministry for Culture and Innovation from the source of the NRDI Fund, Hungary.
\end{acknowledgements}

\bibliographystyle{aa} 
\bibliography{aa} 

\end{document}